\documentclass[abstracton]{scrartcl}
\usepackage[utf8]{inputenc}
\usepackage[english]{babel}
\usepackage{lmodern}

\usepackage{nameref}
\usepackage[colorlinks=true, urlcolor=blue, citecolor=blue, linkcolor=blue]{hyperref}
\usepackage{amsfonts, amsmath, amssymb, amsthm, thmtools}
\usepackage{cleveref}
\usepackage{bm}
\usepackage{graphicx}
\usepackage{algorithm}
\usepackage{algpseudocode}
\usepackage{setspace}
\usepackage{graphicx}
\usepackage{tabularx}
\usepackage{authblk}

\graphicspath{{figures/}{../figures/}}

\usepackage{caption, subcaption}

\usepackage[authoryear]{natbib}
\bibpunct[; ]{(}{)}{,}{a}{,}{;}

\usepackage{lipsum}

\newtheoremstyle{komaplain}
  {\topsep}   % ABOVESPACE
  {\topsep}   % BELOWSPACE
  {\itshape}  % BODYFONT
  {0pt}       % INDENT (empty value is the same as 0pt)
  {\bfseries\sffamily} % HEADFONT
  {.}         % HEADPUNCT
  {5pt plus 1pt minus 1pt} % HEADSPACE
  {}          % CUSTOM-HEAD-SPEC

\theoremstyle{komaplain}

\renewcommand{\epsilon}{\varepsilon}

\title{Identifying regime switches through Bayesian
wavelet estimation: evidence from flood detection in the Taquari River Valley}

\author[1, 2]{Flávia C. Motta}
\author[1]{Michel H. Montoril}
\affil[1]{Department of Statistics, Federal University of São Carlos, São Carlos, SP, Brazil}
\affil[2]{Institute of Mathematics and Computer Sciences, University of São Paulo, São Carlos, SP, Brazil}

\date{May 2023}

\begin{document}

\maketitle

\begin{abstract}
Two-component mixture models have proved to be a powerful tool for modeling heterogeneity in several cluster analysis contexts. However, most methods based on these models assume a constant behavior for the mixture weights, which can be restrictive and unsuitable for some applications. In this paper, we relax this assumption and allow the mixture weights to vary according to the index (e.g., time) to make the model more adaptive to a broader range of data sets. We propose an efficient MCMC algorithm to jointly estimate both component parameters and dynamic weights from their posterior samples. We evaluate the method's performance by running Monte Carlo simulation studies under different scenarios for the dynamic weights. In addition, we apply the algorithm to a time series that records the level reached by a river in southern Brazil. The Taquari River is a water body whose frequent flood inundations have caused various damage to riverside communities. Implementing a dynamic mixture model allows us to properly describe the flood regimes for the areas most affected by these phenomena.
%Two-component mixture models have proved to be a powerful tool for modeling heterogeneity in several cluster analysis contexts. However, most methods based on these models assume a constant behavior for the mixture weights, which can be restrictive and unsuitable for some applications. In this paper, we relax this assumption and allow the mixture weights to vary according to the index (e.g., time) to make the model more adaptive to a broader range of data sets. Similar adaptations in the literature use wavelet bases to estimate the dynamic behavior of the mixture weights but rely on the presumption that the component parameters are known. In this paper, we overcome this limitation and propose an efficient MCMC algorithm to estimate both component parameters and dynamic weights from their posterior samples. We also use wavelet bases due to their widely known advantages when it comes to curve estimation. To illustrate the performance of the proposed estimation, we apply the algorithm to a time series that records the level reached by a river in southern Brazil. The Taquari River is a water body whose frequent flood inundations have caused numerous damage to riverside communities. Implementing a dynamic mixture model allows us to properly describe the flood regimes for the areas most affected by these phenomena.
\end{abstract}

%\tableofcontents

\section{Introduction}

In several data analysis problems, we want to cluster observations between two groups. For instance, in many clinical studies, the goal is to classify patients according to disease absent or present \citep[see][]{Xiao-Hua_mixture_clinical,Rindskopf_mixture_clinical,Hui_mixture_clinical}. In contamination problems found in astronomy investigations, on the other hand, the aim is to separate the objects of interest, called members (e.g., stars), from foreground/background objects contaminating the sample, known as contaminants \citep[see][]{Walker_2009}. In genetics, studies based on microarray data are usually driven to detecting differentially expressed genes under two conditions, e.g., ``healthy tissue \textit{versus} diseased tissue'' \citep[see][]{Bordes_genetics_mixture}. 

%Finite mixture models, contrary to simple parametric models, can offer satisfactory approximations of data sets with multimodal behavior. 
To address these scenarios of bimodal data sets, two-component mixture models have shown to be excellent alternatives to cluster data observations within the group that better describes their features \citep{patra_sen}. In this context, the mixture model with two components will assume that the sample of data observations $y_1, \dots, y_n$ is, in fact, the realization of a random variable $Y$ that belongs to a population composed of two subpopulations, known as mixture components. Thus, at each point $t$, $t=1,\dots, n$, $Y$ is fitted according to some of the mixture components, dictated by a mixture weight $\alpha$. 

This setting may be very restrictive to some data sets. For instance, in epidemiological studies that evaluate the response to medications, the probability of classifying a patient in the group of ``disease present'' must be allowed to vary across time so that the longitudinal effect of the treatment can be properly measured. The same issue arises in quality control problems, where the probability of the supervised system operating in a failure-free regime is also not constant over time. In order to classify those features properly, under a mixture model assumption, the mixture weight should be allowed to vary according to the index (which could be time or location). In other words, it would be appropriate for the mixture weight to present a dynamic behavior.

Assuming dynamic mixture weights for mixture models is an extension that has already been applied in different areas, from traffic flow applications \citep[see][]{Nagy_dynamic_finite_mix} to investigations in genetics \citep[see][]{montoril2019wavelet,michel_helio}. As discussed in \citet{michel_helio}, this generalization is similar to the extension of Hidden Markov Models (HMM) into non-homogeneous Hidden Markov Models (NHMM), first described by \citet{NHMM_hughes}. In both scenarios, one generalizes the model by considering unobserved varying probabilities. In the case of mixture models, those dynamic probabilities are the mixture weights, whereas, in HMM, they are the transition probabilities. It is important to emphasize that, although connected, dynamic mixture weights and transition probabilities are different things.

Considering a ``non-homogeneous'' structure for the mixture model implies that, besides estimating the dynamic mixture weights, one also needs to estimate the component parameters, and that increases the challenge. For instance, in \citet{montoril2019wavelet}, from a frequentist approach, the authors rely on wavelets to perform the estimation of the dynamic weights, where they transform the data in order to deal with a nonparametric heteroscedastic regression. Nonetheless, their procedure depends on assuming known means and variances for the mixture components, which, in practice, may be unrealistic.

In this work, unlike the aforementioned paper, the leading motivation is to provide a Bayesian approach that estimates not only the dynamic mixture weights but also the component parameters of a two-component mixture model. To accomplish this goal, we propose an efficient Gibbs sampling algorithm, which allows the distribution of the posterior draws to be used for inference purposes. Regarding the dynamic mixture weights, we use the data augmentation method by \citet{albert_chib} and incorporate Bayesian wavelet denoising techniques to estimate the dynamic behavior of the mixture weight. We do this to exploit the good properties of wavelets in curves' estimation. 

Wavelets are families of basis functions that can be used to represent other functions, signals, and images as a series of successive approximations \citep{hardle2012wavelets,abramovich2000wavelet}. In statistical applications, these mathematical tools have been successfully used to solve problems in nonparametric regression \citep[see][]{donoho_johnstone_1994ideal,cai1999wavelet}; density estimation \citep[see][]{donoho1993nonlinear,donoho_density1,Hall_patil}; time series analysis \citep[see, e.g.,][]{morettin_96,Priestley_1996,percival_walden_2000}; among many other areas. There is a vast literature that provides a review of wavelets in statistics \citep[see, e.g.,][]{vidakovic1999statistical,ogden1997essential}.
%can be used to represent other functions of interest, such as signals, and images as

In this paper, wavelet bases are applied to enable the estimation of the dynamic mixture weights. To review the mathematical background and the terminology associated with the wavelet theory, in the following section, we provide a short introduction to the wavelet basis functions; the discrete wavelet transform (DWT); and, the Bayesian approach for denoising in a wavelet-based scenario. The remainder of the paper is organized as follows. In Section \ref{sec:model}, we describe the dynamic mixture model considered in this paper and give details related to the MCMC sampling scheme constructed to perform the estimation. In Section \ref{sec:num_ex}, we present some numerical experiments. We first conduct Monte Carlo simulations to evaluate the method in a controlled setting. Then, we apply the MCMC algorithm to a river data set to identify periods when flood inundations occurred.

%\section{Methodology} 
%\subsection{Wavelets}
\section{Wavelets}
In this work, we use the term \textit{wavelets} to refer to a system of orthonormal basis functions for $L_2([0,1])$ or $L_2(\mathbb{R})$. The bases are generated by dyadic translations and dilations of the functions $\varphi(\cdot)$ and $\psi(\cdot)$, known, respectively, as the \textit{scaling} and \textit{wavelet} functions. These systems
of integer-translates and dilates are given by
\begin{align*}
%\label{eq:trans_dila_scaling}
    \varphi_{j_0k}(t) &= 2^{j_0/2}\varphi(2^{j_0}t-k), \quad k \in \mathbb{Z},\\
%    \label{eq:trans_dila_wavelet}  
    \psi_{jk}(t) &= 2^{j/2}\psi(2^j t - k), \quad j, k \in \mathbb{Z}. 
\end{align*}

Thus, for any integer $j_0$ and $J$, a periodic function $f(t) \in L_2([0,1])$ can be approximated in $L_2$-sense as the projection onto a multiresolution space $V_{J}$:
\begin{equation*}
\label{eq:waveletexpansionvj0}
    f(t) = \sum\limits_{k=0}^{2^{j_0}-1}c_{j_0k}\varphi_{j_0k}(t)+\sum\limits_{j=j_0}^{J-1} \sum\limits_{k=0}^{2^j-1} d_{jk}\psi_{jk}(t),
\end{equation*}
where $c_{j_0k}$'s are known as \textit{scaling coefficients} and $d_{jk}$'s are called \textit{detail coefficients}. The former are associated with the coarsest resolution level in witch $f(t)$ was decomposed, $j_0$. As a result, they capture the gross structure of $f(t)$. The detail coefficients, on the other hand, being linked to finer resolution levels, can capture local information about $f(t)$. Put simply, in moving from a coarser resolution level $j$ to a finer $j+1$, we are increasing the resolution at which a function is approximated, thus the expansion coefficients become more descriptive about the local features of $f(t)$.

%DWT
In practice, we access $f(t) \in L_2([0,1])$ through a grid of points in time or space in which $f$ is applied. Therefore,  
%Let $f(t) \in L_2([0,1])$ be the function of interest of some application. In practice, we access $f$ through a grid of points in time or space in which $f$ is applied. Then, 
consider $\bm{f}= (f(1/n),f(2/n), \dots, f(n/n))^T$ to be a vector of samples of $f(t)$ on an equispaced grid of $n$ points, with $n = 2^J$, for some positive integer $J$. To obtain the scaling and detail coefficients that approximate $\bm{f}$, we perform the \textit{discrete wavelet transform} (DWT) of $\bm{f}$. In matrix notation, the DWT of $\bm{f}$ is
\begin{equation}
\label{eq:matrix_W}
    \bm{\theta} = \bm{Wf},
\end{equation}
where $\bm{\theta}=(c_{00},d_{00},\bm{d}_{1}^T, \dots, \bm{d}_{J-1}^T)^T$ is a vector of size $n$, having both scaling and detail coefficients $\bm{d}_j=(d_{j0},d_{j1},\dots, d_{j2^{j}-1})^T$, and $\bm{W}$ is the DWT matrix with $(jk, i)$ entry given by $W_{jk,i}\sqrt{n} \approx \psi_{jk}(i/n)=2^{j/2}\psi(2^ji/n-k)$, $k=0,\dots,2^j-1$, $j= 1, \dots, J-1$. \citep{abramovich1998wavelet}. By orthogonality, the multiplication $\bm{W^T\theta}$ recovers the signal $\bm{f}$. This transformation from wavelet coefficients to fitted values is known as the \textit{inverse discrete wavelet transform} (IDWT). 

One of the main advantages provided by the DWT is the sparse representation generally achieved. As shown by \citet{DONOHO_Unconditional}, wavelets are \textit{unconditional bases} for a range of function spaces, such as Hölder and Sobolev spaces, as well as spaces suitable for representing functions of ‘bounded variation’. As an aside, it is also worth mentioning that using Mallat’s pyramid algorithm \citep{mallat1989theory}, the DWT and IDWT are performed requiring only $\mathcal{O}(n)$ operations, which makes them very efficient in terms of computational speed and storage. These properties help to explain why wavelet bases are excellent tools to address problems of data analysis. In the following section, we present a brief review of handling the denoising problem within the wavelet domain, emphasizing the Bayesian framework due to its central role in the estimation process of this paper.

%An often sought property in wavelet bases is a large number of vanishing moments since it allows sparse representations of functions from a wide set of function spaces, such as Hölder and Sobolev spaces, as well as spaces suitable for representing functions of ‘bounded variation’ \citep{abramovich2000wavelet}. \citet{daubechies1988orthonormal,daubechies1992ten} constructed families of orthonormal wavelet bases whose wavelet functions not only have a finite number of vanishing moments but also are compactly supported. Daubechies wavelets consist of three families: the \textit{extremal-phase}, also known as \textit{daublets}; the \textit{least-asymmetric}, or \textit{symmlets}, and the \textit{coiflets}. %In contrast to the daublets and symmlets that have vanishing moments only for the wavelet function, the coiflets were constructed to also have vanishing moments for the scaling function. 

%\subsubsection{Bayesian wavelet denoising}
\subsection{Bayesian wavelet denoising}
\label{sec:Bayesian_wav_deno}
%Wavelet shrinkage - bayesian

Consider the nonparametric regression model
\begin{equation}\label{eq:nonparametric regression}
    \bm{y}= \bm{f}+\bm{e},
\end{equation}
where $\bm{y}= (y_1, \dots, y_n)^T$ is the vector of observed values, $\bm{f}= (f(1/n), \dots, f(n/n))^T$ is the function of interest applied to a grid of $n$ equally spaced points, and $\bm{e}= (e_1, \dots, e_n)^T$ is a vector of zero-mean random variables. For most applications, $e_t$'s are independent and identically distributed normal random variables with zero mean and constant variance $\sigma^2$. The goal of nonparametric regression is to recover the unknown function $f$ from the noisy observations $\bm{y}$. %unless otherwise specified,

With that in mind, \citet{donoho_johnstone_1994ideal} propose to transform the observations $\bm{y}$ to the wavelet domain, shrink the noisy wavelet coefficients or even equal them to zero, based on some threshold rule, and then estimate $\bm{f}$ by applying the IDWT to the regularized coefficients. This method is known in the literature as \textit{wavelet shinkage}. Therefore, let $n$ be a power of two, $n = 2^J$ for some positive integer $J$. Then, we can represent \eqref{eq:nonparametric regression} in the wavelet domain as
\begin{equation}
    \bm{d}^*=\bm{\theta} + \bm{\epsilon},
\end{equation}
where $\bm{d}^*=\bm{Wy}$, $\bm{\theta} = \bm{Wf}$, and $\bm{\epsilon}= \bm{We}$, with $\bm{W}$ being the DWT matrix.% When using an orthonormal basis to perform the DWT of \eqref{eq:nonparametric regression}, the DWT of $\bm{e}$ is also a vector of independent and identically distributed normal random variables with zero mean and variance $\sigma^2$. %Thus, the DWT of $\bm{y}$ spreads the noise equally over all wavelet coefficients but concentrates most of the signal related to the function $f$ in a few large coefficients \citet{donoho_johnstone_1994ideal}.

From a Bayesian perspective, the wavelet shrinkage technique consists in assigning a prior distribution to each wavelet coefficient of the unknown function. The idea is that, by choosing a prior able to capture the sparseness associated with most wavelet decompositions, we can estimate $\bm{\theta}$, by imposing some Bayes rule on the resulting posterior distribution of the wavelet coefficients. Then, applying the IDWT to the estimated $\bm{\theta}$ gives us an estimation of $\bm{f}$.

One of the most appropriate prior choices for modeling wavelet coefficients are the \textit{spike and slab} priors. First consolidated within Bayesian variable selection methods \citep{spike_slab_1993}, these kinds of prior are a mixture between two components: one that concentrates its mass at values close to zero or even in zero (Dirac delta) and another whose mass is spread over a wide range of possible values for the unknown parameters. Choosing this mixture as prior to the distribution of wavelet coefficients allows the first component, known as \textit{spike}, to capture the null wavelet coefficients, while the second component, called \textit{slab}, describes the coefficients associated with the unknown function. % to be estimated

A  spike and slab prior frequently assigned to wavelet coefficients is the mixture between a point mass at zero and a Gaussian distribution \citep[see, e.g.,][]{abramovich1998wavelet}. In this scenario, each detail wavelet coefficient is distributed following  %In the Bayesian wavelet shrinkage method of \citet{abramovich1998wavelet}, known as \textit{BayesThresh}, the authors propose that %a spike and slab prior given by
\begin{align}
\begin{split}
\label{eq:prior_normal}
    &\pi_{j}\text{N}(0, \upsilon_j^2) + (1-\pi_{j}) \delta_0(\theta_{jk}),\\ 
%    &j=0,\dots, J-1, \quad k = 0, \dots, 2^j -1,
\end{split}
\end{align}%\theta_{jk} \sim 
$k = 0, 1, \dots, 2^j -1$, $j=0, 1, \dots, J-1$, with $\delta_0$ being a point mass at zero. The prior specification is usually completed by assigning a diffuse prior to the scaling coefficient at the coarsest level $c_{00}$. Thus, the sample scaling coefficient obtained from the DWT of the data estimates $c_{00}$ \citep{abramovich1998wavelet}.

%The hyperparameters $\pi_j$ (\textit{sparsity parameter}) and $\upsilon_j^2$ are usually specified for each resolution level $j$ according to four non-negative constants, two determined by the user and two chosen empirically from the data. To complete the prior specification, \citet{abramovich1998wavelet} assigns a diffuse prior to the scaling coefficient at the coarsest level $c_{00}$. Thus, $c_{00}$ is estimated by the sample scaling coefficient obtained from the DWT of the data \citep{abramovich1998wavelet}.

Under the prior \eqref{eq:prior_normal}, the posterior distribution for each detail coefficient is also a mixture between a Gaussian distribution and $\delta_0$, given by
\begin{align}
\label{eq:posterior_normal}
\begin{split}
    \theta_{jk}|d_{jk}^* &\sim \pi_{\text{post}}\text{N}\left(\frac{\upsilon_j^2}{1+\upsilon_j^2}d_{jk}^*, \frac{\upsilon_j^2}{1+\upsilon_j^2}\right) + (1-\pi_\text{{post}}) \delta_0(\theta_{jk}),\\
    \pi_{\text{post}} &= \frac{\pi_j g_{\upsilon_j^2}(d_{jk}^*)}{\pi_j g_{\upsilon_j^2}(d_{jk}^*) + (1-\pi_j) \phi(d_{jk}^*)}, \\
%    j &=0,\dots, J-1, \quad k = 0, \dots, 2^j -1,
    \end{split}
\end{align}
$k = 0, 1, \dots, 2^j -1$, $j=0, 1, \dots, J-1$, where $\phi$ denotes the standard normal density and $g_{\upsilon_j^2}$ denotes the convolution between the slab component in \eqref{eq:prior_normal} (in this case $\text{N}(0, \upsilon_j^2)$) and $\phi$. Using $\gamma$ to denote the slab density and $\star$ to denote the convolution operator, we can write $g= \gamma \star \phi$. It should be stressed that, as shown by \citet{abramovich1998wavelet}, using the posterior medians as the pointwise estimates of $\bm{\theta}$ yields a \textit{thresholding rule}. In other words, we are able to equal the estimated noisy coefficients to zero.

In the \textit{Empirical Bayes thresholding} method by \citet{johnstone2005ebayesthresh,johnstone2005empirical}, the authors propose replacing the Gaussian component in \eqref{eq:prior_normal} with heavy-tailed distributions, such as the Laplace density. This replacement intends to provide larger estimates for the non-null coefficients than those obtained from Gaussian distributions. In this scenario, considering the Laplace density as the slab component, the prior for each detail wavelet coefficient can be written as
%Another Bayesian wavelet denoising method is the \textit{Empirical Bayes thresholding} by \citet{johnstone2005ebayesthresh,johnstone2005empirical}. In this method, the authors propose replacing the Gaussian component that describes the behavior of non-null coefficients with heavy-tailed distributions, such as the Laplace density. This replacement intends to provide larger estimates for the signal coefficients than those obtained from Gaussian distributions. In this scenario, considering the Laplace density as the slab component, the prior for each detail wavelet coefficient can be written as
\begin{align}
\begin{split}
\label{eq:prior_laplace}
    & \pi_{j}\gamma_a(\theta_{jk}) + (1-\pi_{j}) \delta_0(\theta_{jk}), \\
%    & j=0,\dots, J-1, \quad k = 0, \dots, 2^j -1,
\end{split}
\end{align}
$k = 0, 1, \dots, 2^j -1$, $j=0, 1, \dots, J-1$, where $\gamma_a(x)$ denotes the Laplace density with scale parameter $a>0$, i.e.,
\begin{equation}
\label{eq:laplace_density}
   \gamma_a(x) = \frac{a}{2}\exp(-a|x|), \quad x \in \mathbb{R}.
\end{equation}

\citet{johnstone2005ebayesthresh,johnstone2005empirical} thresholding method is called Empirical Bayes because the hyperparameters $\pi_j$ and $a$ are chosen empirically from the data, using a marginal maximum likelihood approach. Thus, for each resolution level $j$ of the wavelet transform, the arguments $\pi_j$ and $a$ that maximize the marginal log-likelihood are selected and plugged back into the prior. Then, the estimation of $\bm{\theta}$ is carried out with either posterior medians, posterior means, or other estimators. Under these circumstances, the posterior distribution is given by
\begin{align}
\label{eq:posterior_laplace1}
\begin{split}
    \theta_{jk}|d_{jk} &\sim \pi_{\text{post}}f_1(\theta_{jk}|d_{jk}) + (1-\pi_{\text{post}}) \delta_0(\theta_{jk}),\\
    \pi_{\text{post}} &= \frac{\pi_j g_a(d_{jk}^*)}{\pi_j g_a(d_{jk}^*) + (1-\pi_j) \phi(d_{jk}^*)}, \\
%    j &=0,\dots, J-1, \quad k = 0, \dots, 2^j -1,
\end{split}
\end{align}
$k = 0, 1, \dots, 2^j -1$, $j=0, 1, \dots, J-1$, with $f_1(\theta_{jk}|d_{jk})$ being the non-null mixture component and $g_a = \gamma_a \star \phi$. It can be shown that $f_1(\theta_{jk}|d_{jk})$ is a mixture of two truncated normal distributions. Define $f_{\text{TN}}(x|\mu, \sigma, \alpha, \beta)$ to be the density of a truncated normal distribution with location parameter $\mu$, scale parameter $\sigma$, minimum value $\alpha$ and maximum value $\beta$. Then, with a slight abuse of notation, we can write $f_1(\theta_{jk}|d_{jk})$ as
\begin{align}
    \label{eq:posterior_laplace2}
    \begin{split}
    f_1(\theta_{jk}|d_{jk}) &= \eta \times f_{\text{TN}}\left(\theta_{jk}\biggl\rvert\frac{d_{jk}}{\sigma_j}-a,1,0,+\infty \right) \\
    &\quad + (1-\eta) \times f_{\text{TN}}\left(\theta_{jk}\biggl\rvert\frac{d_{jk}}{\sigma_j}+a,1,-\infty,0\right),
    \end{split}
\end{align}
where 
\begin{align*}
    \eta = \frac{\exp{(-a\frac{d_{jk}}{\sigma_j})}\Phi(\frac{d_{jk}}{\sigma_j}-a)}{\exp{(a\frac{d_{jk}}{\sigma_j})}\tilde{\Phi}(\frac{d_{jk}}{\sigma_j}+a)+\exp{(-a\frac{d_{jk}}{\sigma_j})}\Phi(\frac{d_{jk}}{\sigma_j}-a)},
\end{align*}
with $\Phi$ denoting the standard normal cumulative function, and $\tilde{\Phi}=1- \Phi$.

%bridge

%\subsection{The model}
\section{The model}
\label{sec:model}

Let $y_1, \dots y_n$ be a random sample from the dynamic Gaussian mixture model
\begin{align}\label{mod:dmix}
\begin{split}
y_t &= (1-z_{t})x_{1t} +  z_{t}x_{2t},\\
x_{kt}| \mu_k, \tau_k^{2} &\sim \text{N}(\mu_k, \tau_k^{-2}), \quad k=1,2,\\
z_t|\alpha_t &\sim \text{Bern}(\alpha_t),\quad t= 1,\dots, n,
%0 \leq \alpha_t &\leq 1, \quad t= 1,\dots, n,
\end{split}
\end{align}
where $z_t$'s are allocation variables that indicate to which mixture component the observations $y_t$'s belong to. The $z_{t}$ have a Bernoulli distribution with parameter $\alpha_{t}$, the mixture weight that has a dynamic behavior. In \eqref{mod:dmix}, the component parameters $\mu_k$ and $\tau_k^2$, $k=1,2$, and the dynamic mixture weights $\alpha_t$, $t=1, \dots, n$, are parameters to be estimated.

Following \citet{albert_chib}, we introduce a data augmentation approach by associating an auxiliary variable $l_t$ to each allocation variable $z_{t}$. In the original work, $l_t = \bm{x}_t^T\bm{\theta} + e_t$ and $e_t \sim \text{N}(0,1)$, where $\bm{{x}_t}$ is a vector of $p$ known covariates and $\bm{\theta}$ is a vector of $p$ unknown parameters. In greater detail, $z_t = 1$, if $l_t > 0$, and $z_t = 0$, otherwise. However, unlike in \citet{albert_chib}, where the design matrix $\bm{X}$ in the probit regression corresponds to the covariates related to $\alpha_t$, in this paper, $\bm{X}=\bm{W}^T$, where $\bm{W}$ is the DWT matrix. Thus, for every $t=1,\dots,n$, we have 
\begin{align}
\begin{split}
\label{eq:lat_def}
l_t &= \bm{x}_t^T\bm{\theta} + e_t,\\
e_t &\sim \text{N}(0,1),\\
\end{split}
\end{align}
where $\bm{x}_t$ corresponds to the $t$-th column of matrix $\bm{W}$ and $\bm{\theta}=(c_{00},d_{00},\bm{d}_{1}^T,\dots, \bm{d}_{J-1}^T)^T$ is the vector of wavelet coefficients, such that $n=p=2^J$. Therefore, the dynamic mixture weight $\alpha_t$, which is the probability of success of $z_t$, is given by the binary regression model,
\begin{equation*}
%    \label{eq:full_con_alpha}
    \alpha_t =\Phi(\bm{x}_t^T\bm{\theta}),
\end{equation*}
where $\Phi$ is the standard Gaussian cumulative function.

\subsection{Bayesian estimation}
%\subsubsection{Bayesian estimation}
\label{sec:bayes_est}

In this paper, the estimation of both component parameters and dynamic mixture weights is performed through a Gibbs sampling algorithm. By giving conjugate prior distributions to the parameters, we sample from their full conditional posterior distributions and make inferences about the parameter values (e.g., point and credible estimates). In this section, we first present the full conditional posterior distributions from which we draw the parameters of \eqref{mod:dmix}. Then, we detail the MCMC algorithm built to perform the sampling.%From a Bayesian framework

In \eqref{mod:dmix}, since we are mostly interested in the estimation of the mixture weights, we assume that the sample $\bm{y}=(y_1,\dots,y_n)^T$ is a time series whose dependence structure is determined by the dynamic behavior of $\alpha_t$'s. In this setting, given the component parameters and the dynamic mixture weights, the observations $y_t$'s are conditionally independent, and we have $p(\bm{y}|\bm{\mu},\bm{\tau^2},\bm{z})=\prod_{t=1}^{n}p(y_t|z_t, \bm{\mu},\bm{\tau^2})$. Thus, the complete-data likelihood function $p(\bm{y}|\bm{\mu},\bm{\tau^2},\bm{z})$ is given by 
\begin{align*}
\prod\limits_{k=1}^{2}\left(\frac{\tau_k^2}{2\pi}\right)^{T_k/2}\exp{\left[-\frac{\tau_k^2}{2}\sum\limits_{t:z_{t}=k-1}(y_t-\mu_k)^2\right]},
\end{align*}
where $T_k = \#\{t:z_{t}=k-1, \, t = 1,2,..., n\}$ and $s_k = \sum\limits_{t:z_{t}=k-1}y_t$  for $k=1,2$. For the complete-data Bayesian estimation of $\bm{\mu}=(\mu_1, \mu_2)^T$ and $\bm{\tau^2}=(\tau^2_1,\tau^2_2)^T$, $p(\bm{y}|\bm{\mu},\bm{\tau^2},\bm{z})$ is combined with prior distributions to obtain the posteriors. A common issue that arises in the Bayesian estimation of mixture models is the invariance of the mixture likelihood function under the relabelling of the mixture components, known as \textit{label switching}. To address this problem in our approach, we adopt the simple constraint $\mu_1 < \mu_2$ and reorder the pairs $(\mu_k, \tau_k^2)$ according to this restriction in the MCMC sampling scheme.

Following the usual practice of assigning independent prior distributions to the component parameters \citep[see][]{Escobar_west_ind,Richardson_Green_ind}, we assume $p(\bm{\mu},\bm{\tau_k^2})=p(\mu_1)p(\tau_1^2)p(\mu_2)p(\tau_2^2)$ and place the following priors on $\mu_k$ and $\tau^2_k$, $k=1,2$, 
\begin{align}
\label{eq:prior_muk}
    \mu_k \sim \text{N}(b_{0k},B_{0k}),\\
    \tau_k^2 \sim \Gamma(c_{0k}, C_{0k}).
\label{eq:prior_tauk}
\end{align}
For the sake of simplicity, hereafter we denote by $[\dots]$ the set of all remaining variables to be considered for the posterior in use. Hence, under the conjugate priors \eqref{eq:prior_muk} and \eqref{eq:prior_tauk}, one obtains the conditional posterior distributions for $\mu_k$ and $\tau_k^2$,
\begin{align}
\label{eq:fullparameterscond1}
    \mu_k|[\dots] \sim \text{N}(b_k, B_k),\\
    \tau_k^2|[\dots] \sim \Gamma(c_k, C_k),
\label{eq:fullparameterscond2}
\end{align}
where 
\begin{align*}
\begin{split}
    B_k &= (B_{0k}^{-1}+\tau_k^2T_k)^{-1},\\
    b_k &= B_k(\tau_k^2 s_k + B_{0k}^{-1}b_{0k}),\\
    \end{split}
    \qquad
    \begin{split}
    C_k &= C_{0k} + \frac{\sum\limits_{t:z_{t}=k-1}(y_t-\mu_k)^2}{2},\\
    c_k &= c_{0k} + \frac{T_k}{2}.
    \end{split}
\end{align*}
%Similarly to $\bm{y}$, 
%$p(\bm{z}|\alpha_1, \dots, \alpha_n)=\prod_{t=1}^{n}p(z_t|\alpha_t)$
It is worth stressing that assuming the mixture weights to have a dynamic behavior does not interfere with the full conditional posteriors of the component parameters, because they are calculated as in the case of the ordinary (static) mixture model.
%As one can see, the fact that the mixture weights are now assumed to have a dynamic behavior, the full conditional posteriors of the component parameters are the same as in the ordinary (static) mixture model. 

Given the observations $\bm{y}$, the component parameters $\bm{\mu}$,  $\bm{\tau^2}$ and $\bm{\alpha}=(\alpha_1,\dots,\alpha_n)^T$, the $z_t$'s are conditionally independent and $p(z_t=1|\bm{y},\bm{\mu},\bm{\tau^2},\bm{\alpha})\propto \alpha_t f_N(y_t|\mu_2,\tau_2^{-2})$. Thus, one can easily show that, for each $t=1,\dots,n$, the full conditional posterior of $z_t$ is given by
\begin{align}\label{eq:fullcondz}
\begin{split}
z_t|[\dots] &\sim \text{Bern}(\beta_t),\\
    \beta_t &= \frac{\alpha_t f_N(y_t|\mu_2,\tau_2^{-2})}{\alpha_t f_N(y_t|\mu_2,\tau_2^{-2}) + (1- \alpha_t)f_N(y_t|\mu_1,\tau_1^{-2})}.
\end{split}
\end{align}

The latent variables introduced in \eqref{eq:lat_def} are unknown. However, given the vector of wavelet coefficients $\bm{\theta}$ and the allocation data $\bm{z}=(z_1, \dots, z_n)^T$, we can use the structure of the MCMC algorithm to draw $l_1, \dots, l_n$ from their posterior distribution, which is
\begin{align}
    \label{eq:full_con_ls}
    \begin{split}
    l_t|[\dots] &\sim \text{N}(\bm{x}_t^T \bm{\theta},1) \text{ truncated at left by 0 if }z_t=1,\\
    l_t|[\dots] &\sim \text{N}(\bm{x}_t^T \bm{\theta},1) \text{ truncated at right by 0 if }z_t=0.
    \end{split}
\end{align}%\bm{z},\bm{\theta}

%With respect to the vector of parameters $\bm{\theta}$, \citet{albert_chib} derived the posterior distribution of $\bm{\theta}$ given $\bm{z}$ and $\bm{l}$ under a diffuse prior. The authors also suggested using a proper conjugate Gaussian distribution for $\bm{\theta}$.

For the vector of parameters $\bm{\theta}$, \citet{albert_chib} derived the posterior distribution of $\bm{\theta}$ given $\bm{z}$ and $\bm{l}$ under diffuse and Gaussian priors. In this work, on the other hand, $\bm{\theta}$ is a vector of wavelet coefficients. As a result, we need a \textit{sparsity inducing} prior able to address the noise $e_t$ in \eqref{eq:lat_def}. Thus, following the discussion in Section \ref{sec:Bayesian_wav_deno}, we suggest using spike and slab priors for the components of vector $\bm{\theta}$. In this scenario, we assume that the entries of $\bm{\theta}$ are mutually independent. For $t=2^{j}+k+1$, $k = 0, \dots, 2^j -1$ and $j=0,\dots, J-1$, this kind of prior can be specified as
  \begin{equation}
\label{eq:prior_mix_generic}
    \theta_t \sim (1-\pi_j)\delta_0(\cdot)+\pi_j \gamma(\cdot),
\end{equation}
where we consider $\gamma$ to be either the Gaussian distribution or the Laplace distribution as presented in \eqref{eq:prior_normal} and in \eqref{eq:prior_laplace}, respectively. Following \citet{abramovich1998wavelet}, the prior specification is completed by assigning a diffuse prior on the scaling coefficient at the coarsest level $c_{00}$, in the first entry of vector $\bm{\theta}$.% Thus, for the mixture priors,  $\theta_1=\bm{w}_1^T \bm{l}$. 

Under \eqref{eq:prior_mix_generic}, the posterior distribution of $\theta_{t}$ is given by %, where $t=2^{j}+k+1$, $k = 0, \dots, 2^j -1$ and $j=0,\dots, J-1$,
    \begin{align}
    \begin{split}
    \label{eq:post_3}
    \theta_{t}|[\dots] &\sim (1-\pi_\text{{post}})\delta_0(\theta_{t}) + \pi_\text{{post}}f_1(\theta_{t}|\bm{w}_t^T\bm{l}),\\
    \pi_\text{{post}}&=\frac{\pi_j g(\bm{w}_t^T \bm{l})}{\pi_j g(\bm{w}_t^T \bm{l}) + (1-\pi_j)\phi(\bm{w}_t^T \bm{l})},
    \end{split}
    \end{align}
where $\bm{w}_t$ is a column-vector corresponding to the $t$-th row of matrix $\bm{W}$, $f_1(\theta_{t}|\bm{w}_t^T\bm{l})$ is the posterior non-null mixture component and $g$ is the convolution between $\gamma$ and the standard normal distribution $\phi$, $g= \gamma \star \phi$.

Regarding the hyperparameters of the spike and slab priors, that is, the sparsity parameter $\pi_j$ and the variance $\upsilon_j^2$ (Gaussian component) or the scale parameter $a$ (Laplace component), we follow the approach in \citet{johnstone2005ebayesthresh,johnstone2005empirical} and estimate them jointly by maximizing the marginal log likelihood function, which is given by %, for $j=0,\dots, J-1$,
\begin{equation*}
    \sum\limits_{i=1+2^j}^{2^{j+1}}\log\{(1-\pi_j)\phi(\bm{w}_i^T \bm{l})+\pi_j g(\bm{w}_i^T \bm{l})\}.
\end{equation*}
These values are then used in \eqref{eq:post_3} to sample the vector $\bm{\theta}$ in the MCMC procedure, which is detailed in Algorithm \ref{alg:gibbs2}.  

%Another alternative would be also assigning priors to $\pi_j$ and $\upsilon_j^2$/$a$ and sample their values from the posteriors. However, this  

\begin{algorithm}
\caption{Gibbs sampling algorithm - Data augmentation}\label{alg:gibbs2}
\begin{algorithmic}[1]
\State Choose number of iterations $N$.
\State Specify initial values for $\bm{\mu}^{(0)}, \,{\bm{\tau^2}}^{(0)}, \, \bm{z}^{(0)}=(z_1^{(0)},\dots,z_n^{(0)})^T$ and $ \bm{\alpha}^{(0)}$.
   \For{$i \gets 1$ to $N$}
      \State Sample $\mu_1^{(i)} \sim p(\mu_1|[\dots])$.  \Comment{See \eqref{eq:fullparameterscond1}}
      \State Sample ${\tau_1^2}^{(i)}\sim p(\tau_1^2|[\dots])$. \Comment{See \eqref{eq:fullparameterscond2}}
      \State Sample $\mu_2^{(i)}\sim p(\mu_2|[\dots])$.  \Comment{See \eqref{eq:fullparameterscond1}}
      \State Sample ${\tau_2^2}^{(i)}\sim p(\tau_2^2|[\dots])$. \Comment{See \eqref{eq:fullparameterscond2}}
      \If{$\mu_2 < \mu_1$} 
      \State Permute the labeling of pairs $(\mu_k^{(i)},{\tau_k^2}^{(i)})$.
      \EndIf
      \State Sample $z_t^{(i)} \sim p(z_t|[\dots])$, for $t=1,\dots,n$. \Comment{See \eqref{eq:fullcondz}}
      \State Sample $l_t^{(i)} \sim p(l_t|[\dots])$, for $t=1,\dots,n$.
      \Comment{See \eqref{eq:full_con_ls}}
      \State Select $\upsilon_j^2\,/\,a$ and $\pi_j$ by marginal maximum likelihood.
      \State Sample $\theta_t^{(i)} \sim p(\theta_t|[\dots])$, for $t=1,\dots,n$. \Comment{See \eqref{eq:post_3}}
      \State Calculate $\bm{\alpha}^{(i)}=\Phi(\bm{W^T\theta})$. \Comment{$\bm{W}$ is the matrix form of the DWT.}
   \EndFor
\end{algorithmic}
\end{algorithm}

As discussed in Section \ref{sec:Bayesian_wav_deno}, using \eqref{eq:prior_mix_generic} as prior for $\theta_t$ allows the posterior medians to act like thresholding rules, equating to zero noisy coefficients. Because of this, we elect the absolute loss as the Bayes rule estimator for the numerical experiments performed using the MCMC method described in Algorithm \ref{alg:gibbs2}.

\section{Numerical Experiments}
\label{sec:num_ex}

In this section, we illustrate the estimation process discussed in the former sections by conducting Monte Carlo experiments and applying it to a river quota data set to identify flood regimes. In both studies, we implement Algorithm \ref{alg:gibbs2} running 6,000 iterations, discarding the first 1,000 as burn-in and performing thinning every 5 draws. We consider the following independent priors for the component parameters: $\mu_1 \sim N(q_1, s^2)$, $\tau_1^2 \sim \Gamma(0.01, 0.01)$, $\mu_2 \sim N(q_3, s^2)$, and $\tau_2^2 \sim \Gamma(0.01, 0.01)$, where $q_1$ and $q_3$ are the first and third quartiles, respectively, of the observed data and $s^2$ is the sample variance. The purpose of using the data statistics is to reduce subjectivity, and, by adopting the quartiles, to segregate the data into two groups.

Concerning the wavelet bases used to perform the transforms, we use the coiflet basis with six vanishing moments. It is important to highlight that, according to other simulated studies, using other Daubechies wavelet bases provides similar results to those achieved by this specific coiflet basis. We do not present these supplementary analyses due to space limitations.

\subsection{Monte Carlo simulations}
\label{sec:monte_carlo}
In our simulated investigations, we generate the artificial data sets by mixing two normally distributed samples of size 1,024, as defined in \eqref{mod:dmix}. In this case, we set the following values for the component parameters: $\mu_1 = 0$, $\mu_2 = 2$, $\tau_1^2=4$ and $\tau_2^2=4$. Concerning the dynamic mixture weights, we employ three different curves for $\alpha_t$: sinusoidal, blocks, and bumps, with the first being defined as $\alpha_t = 0.4\,\cos(2\pi(t+\pi)) + 0.5$, and the last two being rescaled test functions introduced by \citet{donoho_johnstone_1994ideal}.

For all three behaviors of $\alpha_t$, we run 1,000 Monte Carlo replicates. Additionally, we regard both spike and slab priors, discussed in Section \ref{sec:Bayesian_wav_deno}, for the distribution of the wavelet coefficients, namely: the spike and slab prior with Gaussian slab (SSG), and the spike and slab prior with Laplace slab (SSL). Hereafter, we use the acronyms, SSG and SSL, to refer to these priors.

%We implement Algorithm \ref{alg:gibbs2} for each replication of data running 6,000 iterations, discarding the first 1,000 as burn-in and performing thinning every 5 draws. We consider the following independent priors for the component parameters: $\mu_1 \sim N(q_1, s^2)$, $\tau_1^2 \sim \Gamma(0.01, 0.01)$, $\mu_2 \sim N(q_3, s^2)$, and $\tau_2^2 \sim \Gamma(0.01, 0.01)$, where $q_1$ and $q_3$ are the first and third quartiles, respectively, of the observed data and $s^2$ is the sample variance. The purpose of using the data statistics is to reduce subjectivity, and, by adopting the quartiles, to segregate the data into two groups. 

%Concerning the wavelet bases used to perform the transforms, we use the coiflet basis with six vanishing moments. It is important to highlight that we have observed, from other simulated studies, that the use of other Daubechies wavelet bases provides similar results to those achieved by this specific coiflet basis. We do not present these supplementary analyses due to space limitations. 
%Concerning the wavelet bases used to perform the transforms, we use the coiflet basis with six vanishing moments. It is important to highlight that, according to other simulated studies, using other Daubechies wavelet bases provides similar results to those achieved by this specific coiflet basis. We do not present these supplementary analyses due to space limitations.

As mentioned in Section \ref{sec:bayes_est}, the point estimates are the medians of the MCMC chains for each Monte Carlo replicate. To appraise the performance of the estimation as a whole, we calculate the average of these point estimates and their 95\% HPD intervals. The results for the component parameters are presented in \autoref{tab:SSG_par} and \autoref{tab:SSL_par}. It is worth noting that the method, under both priors, performs satisfactorily, with some estimates even coinciding with the parameter values, which, in turn, are encompassed by the HPD intervals in every $\alpha_t$'s scenario.

\begin{figure}[!h]
\centering
\begin{subfigure}[b]{0.5\textwidth}
    \centering
    \includegraphics[width=\textwidth]{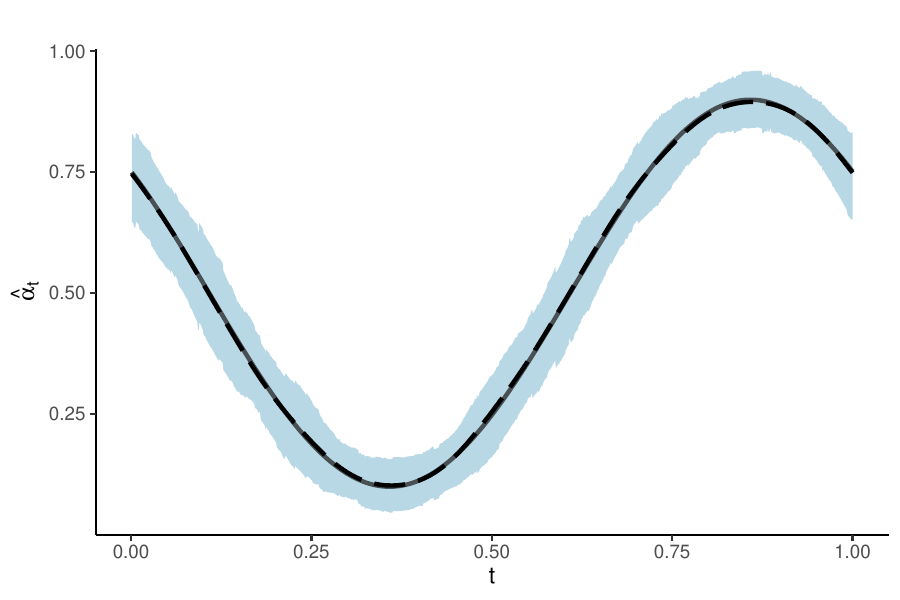}
\end{subfigure}%
\hfill
\begin{subfigure}[b]{0.5\textwidth}
         \centering
         \includegraphics[width=\textwidth]{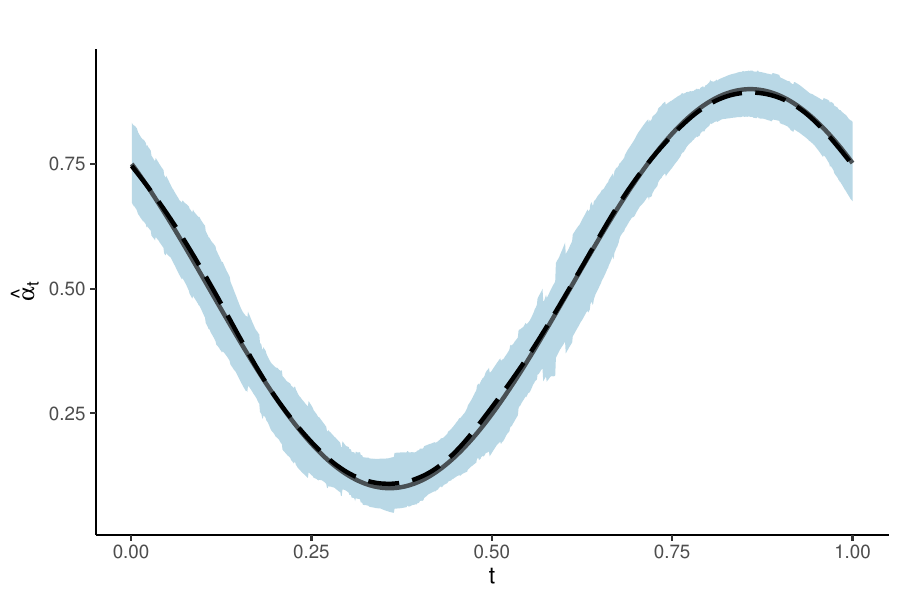}
\end{subfigure}
\begin{subfigure}[b]{0.5\textwidth}
    \centering
    \includegraphics[width=\textwidth]{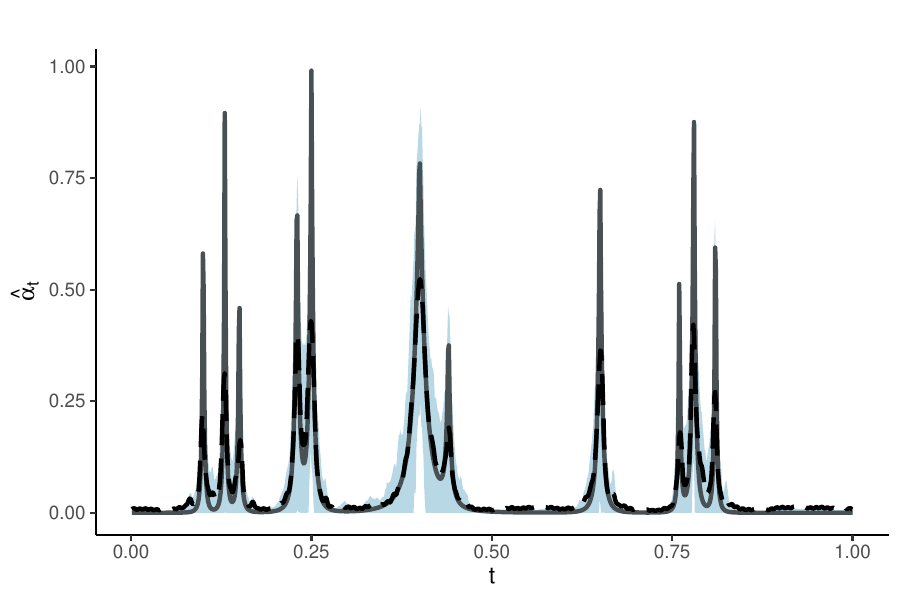}
\end{subfigure}%
\hfill
\begin{subfigure}[b]{0.5\textwidth}
         \centering
         \includegraphics[width=\textwidth]{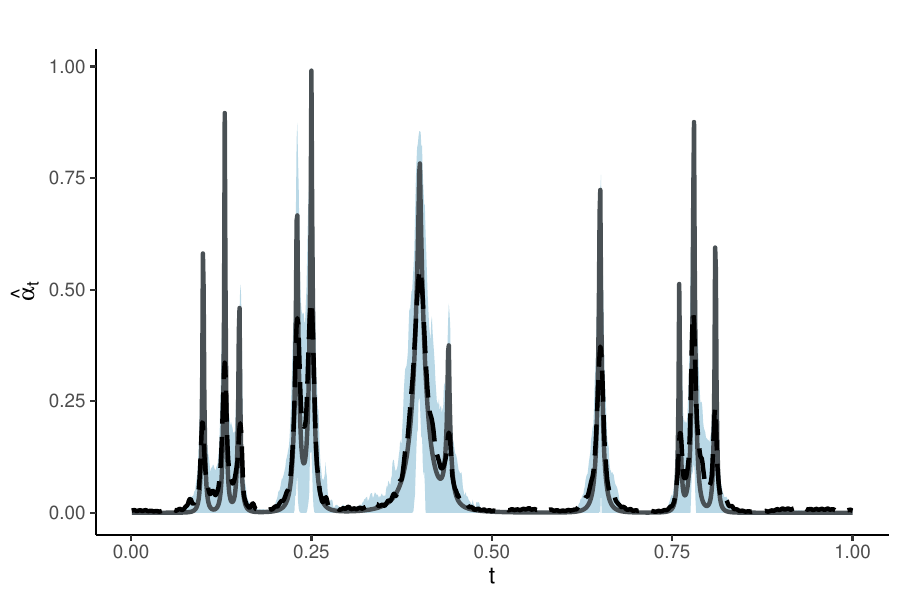}
\end{subfigure}
\begin{subfigure}[b]{0.5\textwidth}
    \centering
    \includegraphics[width=\textwidth]{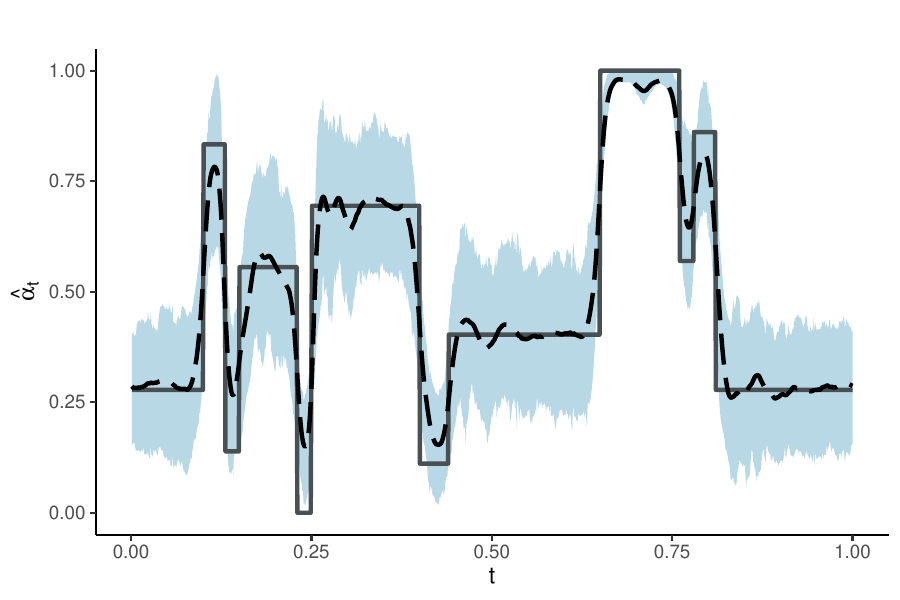}
\end{subfigure}%
\hfill
\begin{subfigure}[b]{0.5\textwidth}
         \centering
         \includegraphics[width=\textwidth]{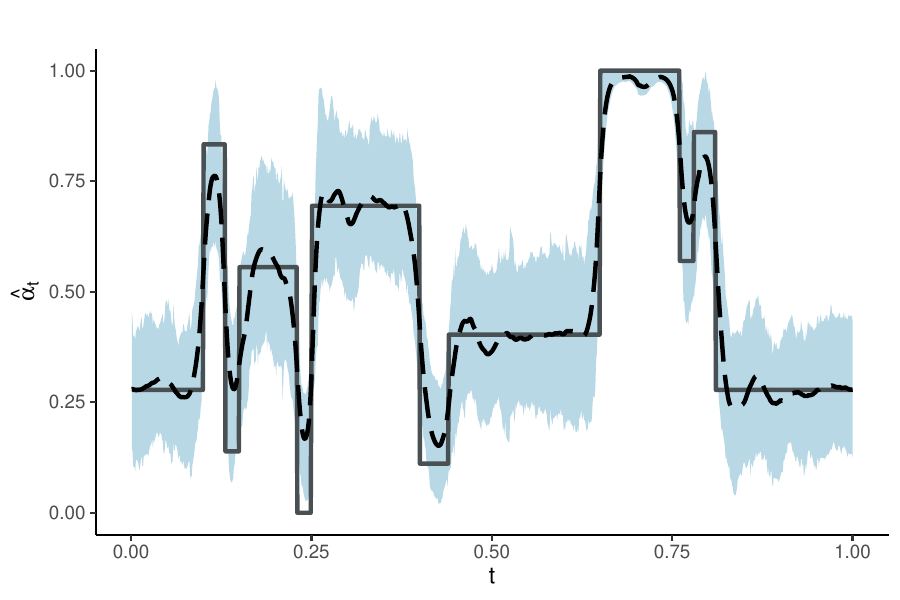}
\end{subfigure}
\caption{Estimates of the $\alpha_t$'s provided by SSG prior (left); and SSL prior (right). The curves assigned to $\alpha_t$ are, respectively: the sinusoidal (top), the bumps (middle), and the blocks (bottom). The full lines correspond to the $\alpha_t$'s curve, the dashed lines correspond to the average of the pointwise estimates and the shaded areas correspond to the 95\% HPD intervals.}
\label{fig:alphas_monte_carlo}
\end{figure}

Regarding the dynamic mixture weights, \autoref{fig:alphas_monte_carlo} shows the results. For the sinusoidal scenario, we see that the method, considering both SSG and SSL priors, succeeds in mimicking the curve's shapes. Although the bumps and blocks functions are less smooth than the sinusoidal, the method still can satisfactorily estimate their curves. In fact, for the bumps, the point estimates not only follow the sharp shape of the function but also captures the null values correctly. For the blocks scenario, the estimates properly mimic the discontinuity regions and the HPD intervals succeed at encompassing the entire curve.

\begin{table}[!h]\small
\caption{Averages of the point estimates (95\% HPD credible intervals) for the component parameters $\mu_1, \tau_1^2, \mu_2$ and $\tau_2^2$, based on 1,000 replications of data sets, considering the SSG prior to $\bm{\theta}$.}
\label{tab:SSG_par}
 \centering
\begin{tabular}{lcccc}
\hline
$\alpha_t$'s curve & $\mu_1$ = 0 & $\tau_1^2$ = 4 & $\mu_2$ = 2 & $\tau_2^2$ = 4 \\ \hline
Sinusoidal & 0.00 (-0.04;0.06) & 4.00 (3.58;4.65) & 2.00 (1.95;2.04) & 4.00 (3.40;4.59) \\
Bumps &  0.00 (-0.04;0.02) & 4.01 (3.59;4.38) & 1.90 (1.60;2.15) & 3.62 (1.06;6.45) \\
Blocks &  0.00 (-0.04;0.06) & 4.06 (3.41;4.71) & 2.00 (1.95;2.06) & 4.00 (3.50;4.63) \\
\hline
\end{tabular}
\end{table}

\begin{table}[!h]\small
\caption{Averages of the point estimates (95\% HPD credible intervals) for the component parameters $\mu_1, \tau_1^2, \mu_2$ and $\tau_2^2$, based on 1,000 replications of data sets, considering the SSL prior to $\bm{\theta}$.}
\label{tab:SSL_par}
 \centering
\begin{tabular}{lcccc}
\hline
$\alpha_t$'s curve & $\mu_1$ = 0 & $\tau_1^2$ = 4 & $\mu_2$ = 2 & $\tau_2^2$ = 4 \\ \hline
Sinusoidal & 0.00 (-0.05;0.05) & 4.05 (3.50;4.62) & 2.00 (1.95;2.04) & 3.99 (3.49;4.50) \\
Bumps &  0.00 (-0.04;0.03) & 3.96 (3.34;4.53) & 1.89 (1.43;2.22) & 3.66 (0.71;6.56) \\
Blocks &  0.02 (-0.15;0.05) & 3.91 (3.40;5.60) & 1.95 (1.28;2.07) & 3.85 (0.82;4.76) \\
\hline
\end{tabular}
\end{table}

%\subsection{aCGH data set}
\subsection{Taquari quota data set}

Part of the Taquari-Antas Hydrographic Basin (TAHB) in the state of Rio Grande do Sul (south of Brazil), the Taquari River is located in the upper domain of the Baixo Taquari-Antas Valley, a region that has been affected by an increasing number of extreme rainfall events in recent decades \citep{Tognoli_Bruski_Araujo_2021}. As a result, on many occasions, the rain excess is not drained efficiently and floods riverside regions. This phenomenon is aggravated in urban areas, where the human occupation of floodplains and the soil impermeability contribute to reducing the infiltration capacity and overloading the drainage system, leading to flood inundations \citep{mastersthesis_roberta}. 

As reported by \citet{de2018caracterizaccao}, Encantado is one of the cities adjacent to the course of the Taquari River most susceptible to fluvial inundations. The geomorphological and topographical characteristics of Encantado's land favor the water accumulation and restrict its drainage \citep{de2018caracterizaccao}. Furthermore, the urbanization of areas with high flood vulnerability in this municipality contributes to intensifying the occurrence of flood inundations \citep{mastersthesis_roberta}. 

\begin{figure}[!h]
    \centering
    \includegraphics[scale=0.75]{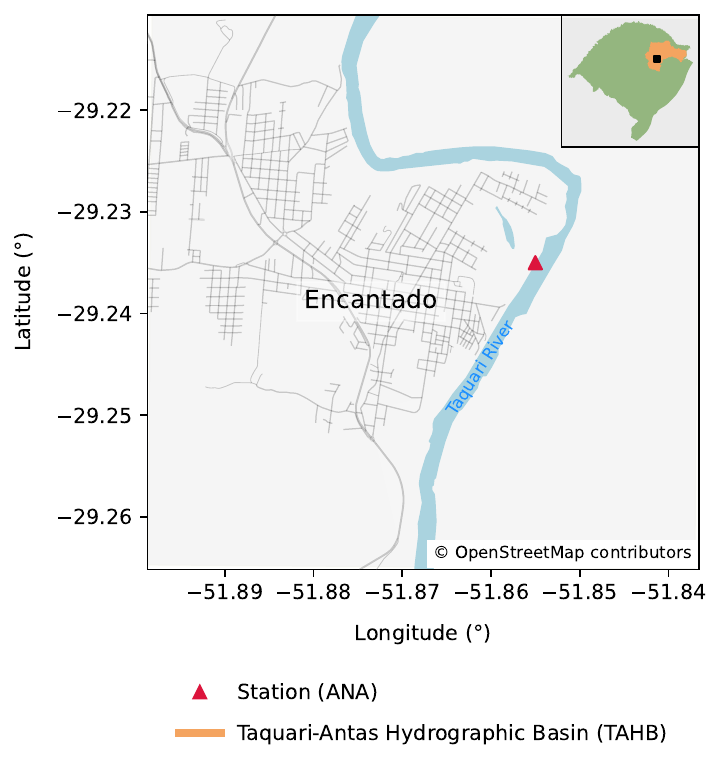}
    \caption{Location map of the fluviometric station in the city of Encantado. In the upper-right corner, the Taquari Antas Hydrographic Basin in Rio Grande do Sul state, south of Brazil.}
    \label{fig:encantado_map}
\end{figure}

Because of these circumstances, we propose implementing Algorithm \ref{alg:gibbs2} to a time series of Taquari's river quota to estimate the probability of an inundation regime in Encantado's urban areas. A river quota is the height of the water body, conventionally measured in centimeters (cm), on a given region of the riverbank. The data set corresponds to the records of Encantado´s fluviometric station identified by the code 86720000. The monthly time series of this station comes from the Hidroweb system, an integrated platform of the National Water Resources Management System (SINGREH) available at \url{https://www.snirh.gov.br/hidroweb/serieshistoricas}. \autoref{fig:encantado_map} shows a map of Encantado, highlighting the station used in this study. 

\begin{figure}[!h]
    \centering
    \includegraphics[scale=0.65]{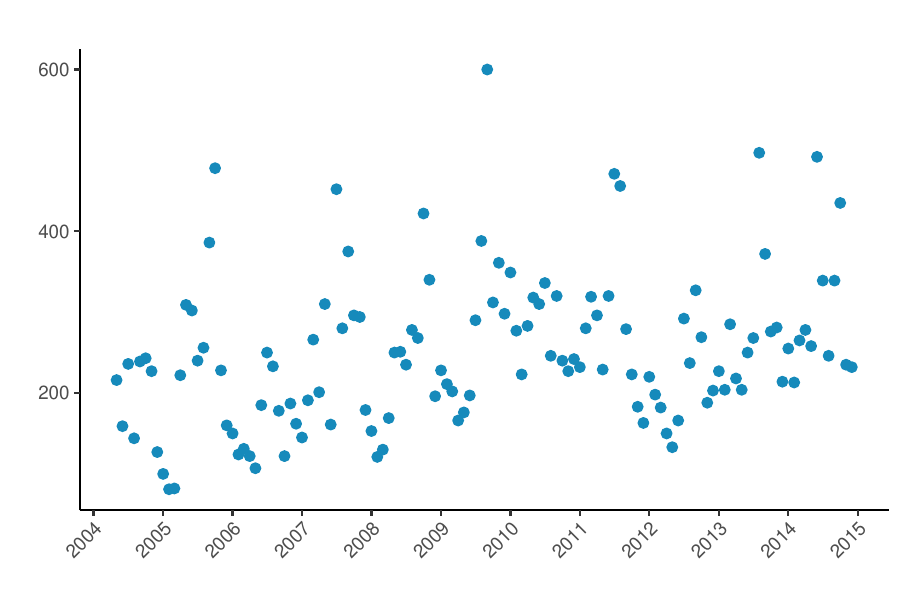}
    \caption{Monthly average of Taquari's river quota (cm) from May 2004 to December 2014.}
    \label{fig:river_quote_plot}
\end{figure}

To validate the estimated probabilities, we use a report from the Brazilian Geological Survey (CPRM) \citep{Peixo_Debo_rep} that records the months when floods occurred in Encantado. Therefore, we can see if the estimates of the mixture weight properly describe the flood regimes, \textit{no inundation} and \textit{inundation}, for each month. It is worth highlighting that since inundations can last for a couple of days or even more, there are no records of the specific days when these events took place, only the months. Because of that, and considering that the model is a mixture of two Gaussian distributions, we use the monthly average of the Taquari quota to estimate the probability associated with flood inundations. The period analyzed was from May 2004 to December 2014, consisting of 128 observations. \autoref{fig:river_quote_plot} presents this data set.

%We are also interested in estimating the parameters of the component distributions that describe each flood regime. \autoref{tab:taquari.par} shows these estimates.  
%
%Algorithm \ref{alg:gibbs2} is implemented running N = 6,000 iterations with burn-in B = 1,000 and lags of L = 5. As in \autoref{sec:monte_carlo}, we also consider the following independent priors for the component parameters: $\mu_1 \sim N(q_1, s^2)$, $\tau_1^2 \sim \Gamma(0.01, 0.01)$, $\mu_2 \sim N(q_3, s^2)$, and $\tau_2^2 \sim \Gamma(0.01, 0.01)$, where $q_1$ and $q_3$ are the first and third quartiles of the observed data and $s^2$ is the sample variance. 

\begin{figure}[!h]
\centering
\begin{subfigure}[b]{0.5\textwidth}
    \centering
    \includegraphics[width=\textwidth]{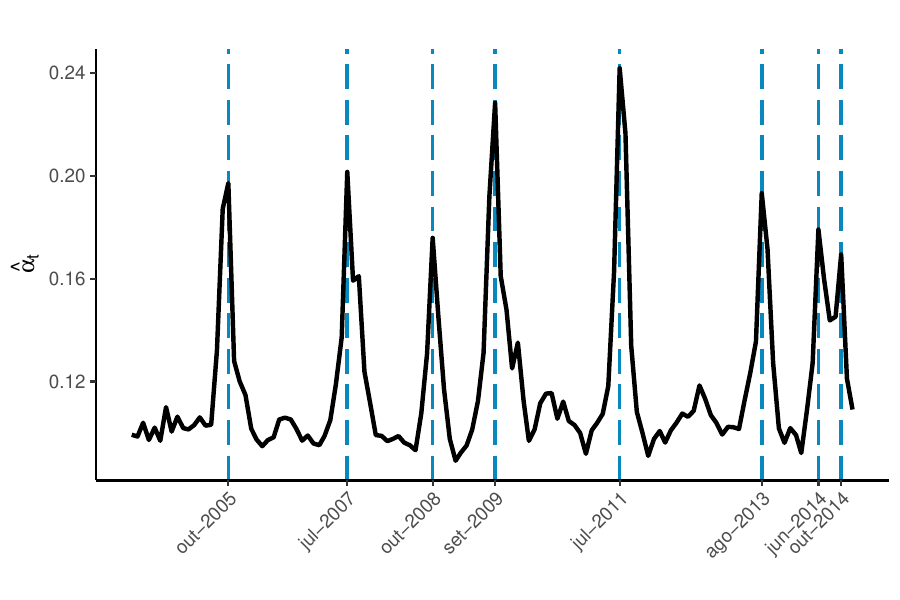}
  \label{fig:subb1}
\end{subfigure}%
\hfill
\begin{subfigure}[b]{0.5\textwidth}
         \centering
         \includegraphics[width=\textwidth]{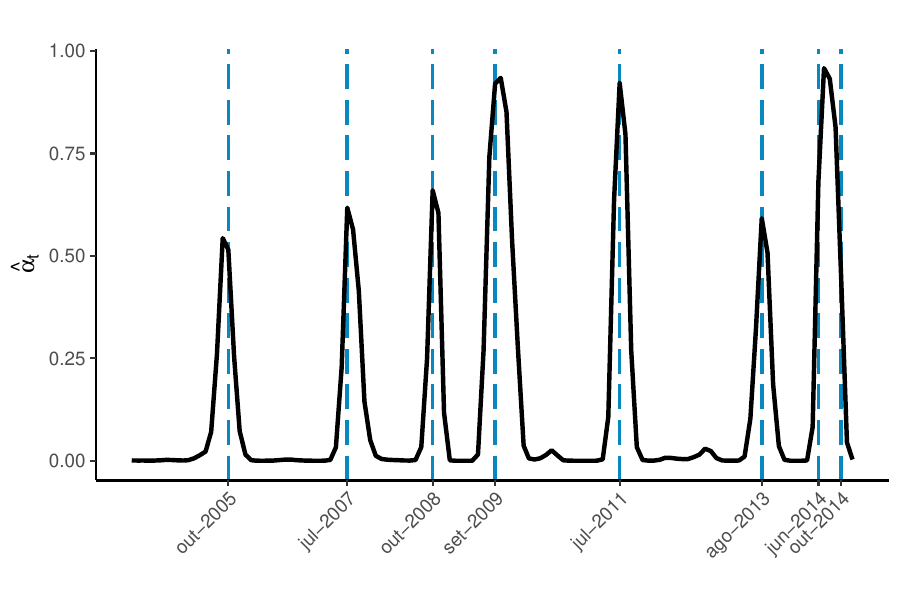}
  \label{fig:subb2}
\end{subfigure}
\caption{Estimates of the $\alpha_t$'s of the Taquari quota data provided by SSG prior (left); and SSL prior (right). The full (black) lines correspond to the point estimates (medians) and the dashed (blue) lines mark the months when flood inundations were reported by \citet{Peixo_Debo_rep}.}
\label{fig:alphas_taquari}
\end{figure}

\autoref{tab:taquari.par} shows the point estimates for the component parameters that describe each flood regime. Note that the results provided by the method under the SSG prior are similar to those achieved by it assigning the SSL prior to the distribution of wavelet coefficients. Concerning the dynamic mixture weights, \autoref{fig:alphas_taquari} shows the estimates considering both priors for $\bm{\theta}$. By analyzing the results, we see that using the SSL prior allows estimating higher peaks for the probabilities related to inundation periods than using the SSG prior. In fact, under a Bayes classifier, if the method employs the SSG prior, it can detect neither the months when flood episodes were reported nor change points ($\{t:\alpha_t=0.5\}$). 

%However, if we consider the period from June 2014 to October 2014, we see that the method under the SSL cannot indicate lower probabilities for months in which floods did not occur (July, August, and September). Conversely, when using the SSG prior, the estimates during these months are lower than those for the months when floods did occur.%in inudantions periods
%utilizando um classificador de Bayes, por exemplo, ele não conseguiria detectar períodos de enchente nem pontos de mudança ($\{t:\alpha_t=0.5\}$).

In summary, the method provides results consistent with the data on flood inundations in Encantado available in other works and reports \citep[see][]{Peixo_Debo_rep,Tognoli_Bruski_Araujo_2021}. In addition, choosing the Laplace density in the spike and slab prior tends to provide dynamic weight estimates more capable of detecting floods. %both spike and slab priors considered have shown to be good choices for describing the distribution of wavelet coefficients. Nevertheless, choosing the Laplace density to model the non-null coefficients offers larger values for the coefficients that are information instead of noise, at least for this data set.  

%Apart from estimating the probability 
%In general, both priors provides similar estimates for  

%history of flood-related disasters, with recurring public and private losses. Between 1980 and 2007, 55% of flood events resulted in a Municipal Emergency Situation decree, with significant economic losses   

%\begin{table}[!h]\small
%\caption{Medians (95\% HPD credible intervals) for the component parameters $\mu_1, \tau_1^2, \mu_2$ and $\tau_2^2$ of the Taquari quota data set, based on the MCMC samples.}
%\label{tab:taquari.par2}
% \centering
%\begin{tabular}{lcccc}
%\hline
%Prior to $\bm{\theta}$ & $\mu_1$ & $\tau_1^2$ & $\mu_2$ & $\tau_2^2$ \\ \hline
%SSG prior & 227.07 (210.09; 242.89) & 2.30e-4 (1.54e-4; 3.15e-4) & 405.01 (316.38; 483.35) & 1.14e-4 (2.56e-5; 3.42e-4) \\
%SSL prior & 0.23 (0.16; 0.30) & 5.18 (3.97; 6.51) & 3.61 (2.74; 4.55) & 0.26 (0.13; 0.44) \\
%\hline
%\end{tabular}
%\end{table}

\begin{table}[!h]\small
\caption{Medians (95\% HPD credible intervals) for the component parameters $\mu_1, \tau_1^2, \mu_2$ and $\tau_2^2$ of the Taquari quota data set, based on the MCMC samples.}
\label{tab:taquari.par}
 \centering
\begin{tabular}{lcc}
\hline
Parameters & SSG prior & SSL prior \\ \hline
$\mu_1$    & 227.07 (210.09; 242.89)    & 220.60 (206.25; 236.28)\\
$\tau_1^2$ & 2.30e-4 (1.54e-4; 3.15e-4) & 2.58e-4 (1.77e-4; 3.45e-4)\\
$\mu_2$    & 405.01 (316.38; 483.35)    & 400.20 (355.72; 439.54)\\
$\tau_2^2$ & 1.14e-4 (2.56e-5; 3.42e-4) & 1.04e-4 (3.65e-5; 1.85e-4)\\
\hline
\end{tabular}
\end{table}

%the shaded areas correspond to the 95\% HPD intervals.

\section{Conclusion}

This paper presents an approach to identify regime switches in bimodal data sets. We use a two-component mixture model whose mixture weight varies according to some index, like time. This adaptation makes the model more flexible and adaptive to a broader range of clustering and classification problems. Furthermore, we use wavelet bases to estimate the dynamic behavior of the mixture weight due to their excellent properties when it comes to curves' estimation. However, unlike other approaches in the literature that also rely on wavelets  \citep[see][]{montoril2019wavelet}, here we consider a Bayesian framework and propose estimating the dynamic weights and the component parameters jointly through an efficient Gibbs sampling algorithm.

We analyze the performance of this MCMC algorithm by conducting Monte Carlo experiments and illustrate the approach with an application to a river quota data set. Results from the simulations show that the method provides good estimates for the component parameters and the dynamic weights even when the function behind $\alpha_t$'s behavior is rougher. Additionally, the estimation performance using SSG prior is similar to the performance achieved when SSL prior is employed. The same does not apply to the results obtained in the river quota data set. For this application, we notice that implementing the method under the SSG prior to the wavelet coefficients yields smaller values for the probabilities associated with inundations occurrence than the estimates provided by using the SSL prior. This is likely because the Gaussian distribution does not have heavy tails, unlike the Laplace distribution.

%Nonetheless, in both scenarios, the method indicates higher probabilities (peaks) related to the months when inundation episodes were reported.

%SSG resultados analogos a SSL, na aplicação tem dificuldade de classificar grupos diferentes (Conclusão)- isso se dá provavelmente (likely) porque a distribuição gaussiana não tem caudas pesadas ao contrário da laplace

\bibliographystyle{apalike}
\bibliography{bibliography.bib}

\end{document}